\newcommand{\araa}{ARA\&A} 
\newcommand{\aap}{A\&A}
\newcommand{\apj}{ApJ} 

\newcommand{\apjl}{ApJ}
\newcommand{\mnras}{MNRAS}

\newcommand{\nat}{Nature}
\newcommand{\physrep}{Phys. Rep.}

\newcommand{\na}{New Ast.}

\documentclass{iau} 
\usepackage{graphicx}
\usepackage{natbib}
\usepackage{epstopdf}

\title[Jet Structure and Dynamics from GW-EM Counterparts] 
{Revealing Short GRB Jet Structure and Dynamics with Gravitational Wave Electromagnetic Counterparts}

\author[G P Lamb \& S Kobayashi]   
{Gavin P Lamb$^1$
 \and Shiho Kobayashi$^1$}

\affiliation{$^1$Astrophysics Research Institute, Liverpool John Moores University, Liverpool, L3 5RF, UK \\ email: {\tt g.p.lamb@2010.ljmu.ac.uk}}

\pubyear{2017}
\volume{338}  
\jname{Gravitational Wave Astrophysics: Early Results from GW Searches and Electromagnetic Counterparts}
\editors{Gabriela Gonz\'alez, eds.}
\begin{document}

\maketitle

\begin{abstract}
Compact object mergers are promising candidates for the progenitor system of short gamma-ray bursts (GRBs). Using gravitational wave (GW) triggers to identify a merger, any electromagnetic (EM) counterparts from the jet can be used to constrain the dynamics and structure of short GRB jets. GW triggered searches could reveal a hidden population of optical transients associated with the short-lived jets from the merger object. If the population of merger-jets is dominated by low-Lorentz-factors, then a GW triggered search will reveal the on-axis orphan afterglows from these failed GRBs. By considering the EM counterparts from a jet, with or without the prompt GRB, the jet structure and dynamics can be constrained. By modelling the afterglow of various jet structures with viewing angle, we provide observable predictions for the on- and off- axis EM jet counterparts. The predictions provide an indication for the various features expected from the proposed jet structure models.
\end{abstract}

\firstsection 
\section{Introduction}

Gravitational wave (GW) driven mergers of neutron star (NS) or black hole (BH) - neutron star binary systems are the most promising candidate for the progenitor of short duration $\gamma$-ray bursts (GRBs) \citep[e.g.][]{1989Natur.340..126E,1992ApJ...395L..83N, 1993Natur.361..236M,2007ARep...51..308B,2007PhR...442..166N,2014ARA&A..52...43B}.
Mergering NS/BH-NS systems will be detectable by gravitational wave observatories where the mergers occur within the detection horizon \citep{2013ApJ...767..124N}.
Amongst the electromagnetic (EM) counterparts are isotropic macro/kilo-nova \citep[e.g.][]{1998ApJ...507L..59L, 2013ApJ...775...18B, 2013ApJ...775..113T, 2013MNRAS.430.2121P, 2014MNRAS.441.3444M, 2014ApJ...780...31T, 2017arXiv170809101T, 2016ApJ...829..110B, 2016AdAst2016E...8T, 2017LRR....20....3M}, radio counterparts \citep[e.g.][]{2011Natur.478...82N, 2014MNRAS.437L...6K, 2015MNRAS.452.3419M, 2015MNRAS.450.1430H, 2016ApJ...831..190H}, wide angle coccoon emission \citep{2016arXiv161001157L, 2017ApJ...834...28N, 2017arXiv170510797G, 2017arXiv171100243K}, resonant shattering, merger-shock or precursor flares \citep{2013AAS...22134601T, 2014MNRAS.437L...6K, 2016MNRAS.461.4435M, 2017arXiv171103112S}, GRBs \citep[e.g.][]{2014MNRAS.445.3575C, 2016A&A...594A..84G, 2017arXiv170807488K, 2017arXiv170807008J}, failed GRBs (fGRB) \citep{2000ApJ...537..785D, 2002MNRAS.332..735H, 2003NewA....8..141N, 2003ApJ...591.1097R, 2016ApJ...829..112L, 2017IAUS..324...66L, 2017MNRAS.472.4953L}, and off-axis orphan afterglows \citep[e.g.][]{2002MNRAS.332..945R, 2002ApJ...570L..61G, 2007A&A...461..115Z, 2013ApJ...763L..22Z, 2016arXiv161001157L, 2017ApJ...835....7S}.

GRBs are the dissipation of energy within relativistic jets launched due to the rapid accretion of material following the formation of a stellar mass BH (or a NS).
Relativistic jets from accreting BH systems are seen on all mass scales, from the stellar mass BHs in some X-ray binaries and GRBs, to the super-massive BHs at the centres of galaxies.
The universality of these jets indicates that the physical processes involved in the formation, collimation and acceleration are likely to be similar.
These similarities have given rise to a number of attempts to show scaling relations for the observables from these systems \citep[e.g][]{2003MNRAS.345.1057M, 2003MNRAS.343L..59H, 2004A&A...414..895F, 2004A&A...415L..35T, 2012ApJ...757...56Y, 2012Sci...338.1445N, 2014ApJ...780L..14M, 2017MNRAS.472..475L, 2017MNRAS.470.1101W}.
For the transient jets associated with GRBs, external shocks form as the jet decelerates leading to an afterglow \citep{1998ApJ...497L..17S}.

The highly variable non-thermal emission of GRBs can only be explained if the jet or outflow is ultrarelativistic, with typical bulk Lorentz factors $\Gamma\sim100$ \citep[][etc.]{2002ARA&A..40..137M}.
As the jet decelerates it produces a broadband afterglow, this emission is beamed within an angle $1/\Gamma$.
When $1/\Gamma > \theta_j$, the jets half-opening angle, a break in the afterglow lightcurve will be seen.
For short GRBs, where the arrival time duration for $90\%$ of the prompt emission is $<2$s, the opening angle of the jet is poorly constrained.
The range of opening angles from short GRB afterglow break time measurements indicates a wide range of jet half-opening angles, $2^\circ$ at the narrowest and $\geq 25^\circ$ for the widest \citep{2015ApJ...815..102F}.
Energy distribution (or structure) within the jet is usually assumed to be homogeneous.

\begin{figure}
\centering
\includegraphics[width=\textwidth]{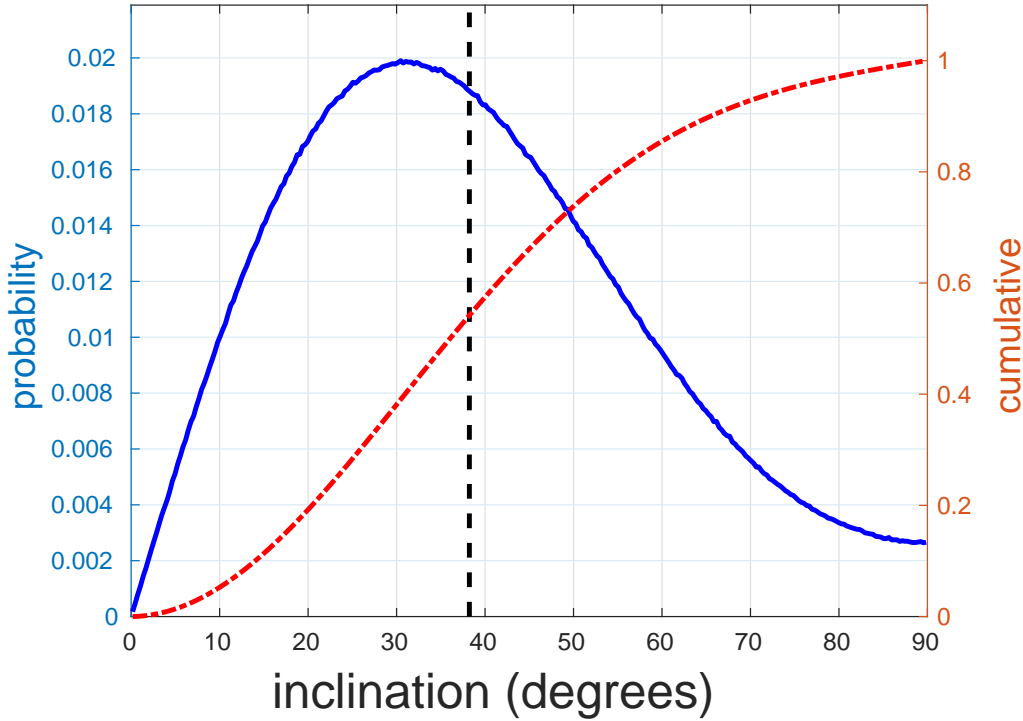}
\caption{By considering the GW strain from a merger as a function of inclination, the distribution of system inclinations can be determined. For all GW
detected mergers at a fraction of the maximum detectable luminosity distance, the probability of a system being inclined at given angle is shown with the blue
solid line. The mean system inclination for this distribution is the dashed black line. The red dashed–dotted line is the cumulative distribution. Figure from Lamb \& Kobayashi 2017c}
\label{GWfig}
\end{figure}

As jets are collimated, the probabilty that a system with a bi-polar outflow is inclined towards an observer is $\sim \theta_j^2/2$.
For an isotropic distribution of systems, the on-axis probability is small.
However, for a GW detected system where GWs are strongest in the polar directions, there is a GW Malmquist bias towards merging systems with a low-inclination \citep{1993ApJ...417L..17K,2010ApJ...725..496N,2011CQGra..28l5023S}.
Figure \ref{GWfig} shows the probability for a GW detected system with a given inclination, and the cumlative fraction of events inclined within an inclination angle.
For jets with a wide half-opening angle, or where the EM counterparts from the jet are bright at inclinations $20-40^\circ$ e.g. structured jets, then the potential to observe a jet origin EM counterpart following a GW detected NS/BH-NS merger is reasonable, $\sim 0.2 - 0.6$ \citep{2017MNRAS.472.4953L}.

In \S \ref{Low} we discuss the expected excess in a population of merger-jets where the dominant fraction has a low-$\Gamma$.
In \S \ref{Structure} we describe the lightcurves from the afterglow of various structured jets and give an indication of how some structure models can enhance the fraction of bright optical transients associated with GW detected NS/BH-NS mergers.
\S \ref{Sum} gives a summary of \S \ref{Low} and \S \ref{Structure} plus additional comments and discussion regarding a population of low-$\Gamma$ or structured jets.

\section{Low-$\Gamma$ Jets - failed GRBs}\label{Low}

Where baryon loading of a relativistic outflow is efficient the bulk Lorentz-factor of the jet will be low i.e. $\Gamma<<40$ \citep[e.g.][]{2013ApJ...765..125L}, where $\Gamma$ is very low the GRB will be suppressed resulting in a failed-GRB and an afterglow-like transient. 
Failed GRBs, or dirty fireballs, were first discussed by \cite{2000ApJ...537..785D,2002MNRAS.332..735H,2003NewA....8..141N} and \cite{2003ApJ...591.1097R} as potential sources of X-ray, optical and radio transients similar in appearance to a GRB afterglow but without the prompt high-energy trigger \citep[e.g.][]{2013ApJ...769..130C}.
A population of low-$\Gamma$ jets from mergers may go undetected, GW astronomy provides a trigger to search for such failed-GRB transients from merger-jets \citep{2016ApJ...829..112L}.

Relativistic outflows become optically thin at the photospheric radius, $R_p\propto (E/\Gamma)^{1/2}$, and the minimum variability timescale for the prompt $\gamma$-ray emission constrains the radius from which these high-energy photons are emitted, $R_d \sim c\delta t\Gamma^2$.
For a bright GRB the dissipation radius should be above the photospheric radius, $R_d>R_p$, and $\Gamma \gtrsim 80 E_{51}^{1/5} \delta t_{-1}^{-2/5}$ where $E_{51}= E/(10^{51} {\rm ~erg})$ and $\delta t_{-1}= \delta t/(0.1 {\rm ~s})$.
If $R_d<<R_p$ then the $\gamma$-rays will be injected into an optically thick plasma.
The high-energy photons will be coupled to the plasma and adiabatically cool until the optical depth reaches unity.
Additionally the photons will undergo pair-production and Compton down-scattering that progressively thermalizes the distribution \citep{2005ApJ...635..476P,2007ApJ...666.1012T,2014ApJ...782....5H}.

To successfully produce a GRB at typical short GRB jet kinetic energies and efficiencies, the bulk Lorentz factor for an outflow should be $\Gamma\gtrsim20-30$.
For jets where the Lorentz-factor is below this limit, the outflow will not produce a GRB but will still result in a broadband afterglow that could be detected as an on-axis orphan afterglow.
If we consider a cosmological population of merger jets that follow a \cite{2015MNRAS.448.3026W} redshift and initial luminosity function, but where the bulk Lorentz-factor follows a distribution $N(\Gamma) \propto \Gamma^{-7/4}$, then $\sim 91\%$ of the population result in failed GRBs or GRBs too faint to be {\it Swift}/BAT detectable.
Where the volume is limited to $z\leq0.07$, approximately the face-on limit for NS-NS mergers detectable by advanced LIGO \citep[][etc.]{2016LRR....19....1A} and there is an associated GW detection, then the fraction is $\sim 78\%$ failed GRBs \citep{2016ApJ...829..112L}.
These fractions only consider a population of systems with inclinations less than a jets half-opening angle.

\begin{figure}
\centering
\includegraphics[width=\textwidth]{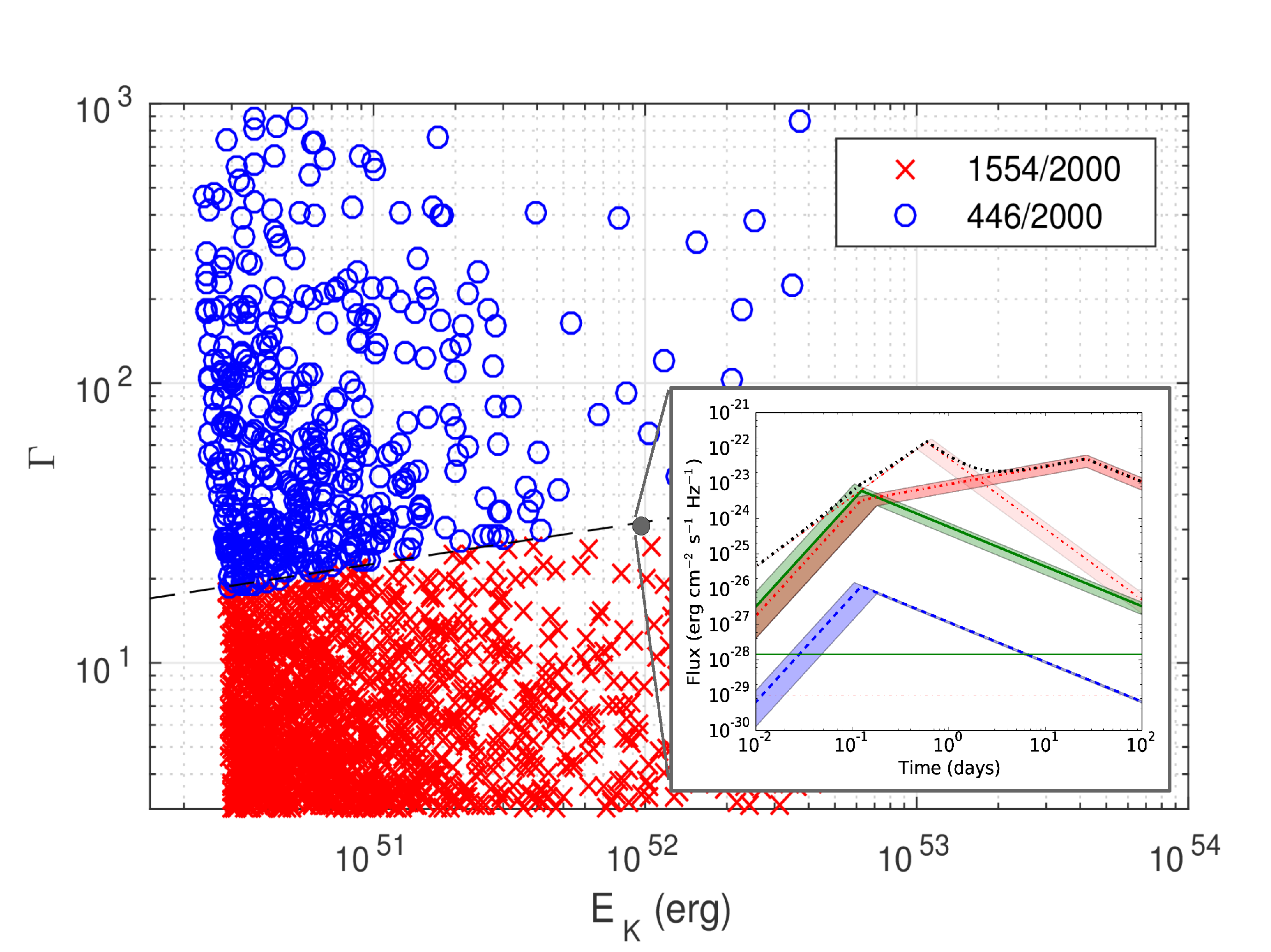}
\caption{A Monte Carlo population of merger-jets oriented towards an observer within $z\leq0.07$. Isotropic equivalent jet kinetic energy is shown on the $x$-axis, whilst the jet initial bulk Lorentz factor is shown on the $y$-axis. Blue circle indicates a {\it Swift}/BAT detectable short GRB, a red cross indicates a failed GRB. Inset: lightcurve for an on-axis low-luminosity GRB at 40 Mpc with parameters indicated by the grey circle. Red dashed-dotted line is forward- and reverse- shock emission at 10 GHz; green solid line is forward-shock $r$-band emission; and blue dashed line is forward-shock X-ray emission. Horizontal lines indicate flux level: 1$\mu$Jy at 10 GHz, red dashed-dotted; and magnitude 21 at $r$-band, green solid line. The lower limit of the $y$-axis indicates the X-ray sensitivity of $\sim 0.4$ $\mu$Crab at 4 keV. Figure adapted from Lamb \& Kobayashi 2016, 2017d}
\label{fig1}
\end{figure}

The distribution of GRBs and failed GRBs with jet kinetic energy and bulk Lorentz-factor are shown in Figure \ref{fig1} for the local events, $z\leq0.07$.
The inset shows an example afterglow lightcurve for a merger-jet at 40 Mpc, with a $\Gamma=30$ and a jet kinetic energy $E_{\rm k}\sim 10^{52}$ erg in an ambient medium with particle number density $n=(9\pm6)\times 10^{-3}$ cm$^{-3}$, the shaded regions represent the density uncertainty;
such a merger-jet would produce a very low-luminosity GRB just below the detection limit for {\it Swift}/BAT.
The afterglow at radio (10 GHz), optical ($r$-band) and X-ray ($10^{18}$ Hz) are shown.
The afterglow is bright for an on-axis observer \citep{2017arXiv171005857L}.

For on-axis orphan afterglows at a distance less than $\sim 300$ Mpc, the optical peak flux is brighter than magnitude 21 for $\sim 85\%$ of cases whilst 10 GHz and X-ray are always brighter than the detection limits for the various facilities at these wavelengths \citep{2016ApJ...829..112L}, e.g. VLA and XRT respectively.
Optical and X-ray emission peaks on the same timescale, typically 0.1-10 days after a GW merger signal.
Radio emission will peak from 10 days after the merger.
Jets with a lower Lorentz-factor will peak later and be fainter due to the characteristic frequency being below that of the observation frequency at the peak time.

\section{Jet Structure}\label{Structure}

Structured jets have a jet energy per steradian (or  other parameter) that varies with angle from the central axis \citep[e.g.][]{2002ApJ...570L..61G,2002ApJ...571..876Z,2002MNRAS.332..945R,2004MNRAS.354...86R, 2003ApJ...594L..23V,2003A&A...400..415W,2003ApJ...591.1075K,2005MNRAS.363.1409P, 2005ApJ...626..966P,2007ApJ...656L..57J,2011MNRAS.417.2161B}.
A relativistic jet may have an intrinsic structure due to the formation mechanism \citep[e.g.][]{2003ApJ...584..937V,2003ApJ...594L..23V} or as the jet breaks out of the medium immediately around the central-engine \citep[e.g.][]{2002bjgr.conf..146L, 2003ApJ...594L..19L, 2003ApJ...586..356Z, 2004ApJ...608..365Z, 2005ApJ...629..903L, 2010ApJ...723..267M, 2015MNRAS.447.1911P}.
As a jet breaks out from the dynamical ejecta associated with a NS/BH-NS merger, the jet will lose the collimating pressure of a cocoon \citep{2011ApJ...740..100B}.
This may result in the wider components of the jet having a lower energy or Lorentz-factor distribution with angle from the jet-core region;
such jet structures have been proposed as potential EM counterparts to GW detected NS/BH-NS mergers \citep[e.g.][]{2017MNRAS.472.4953L,2017arXiv170807488K,2017arXiv170807008J,2017arXiv171000275X}.

\begin{figure}
\includegraphics[width=\textwidth]{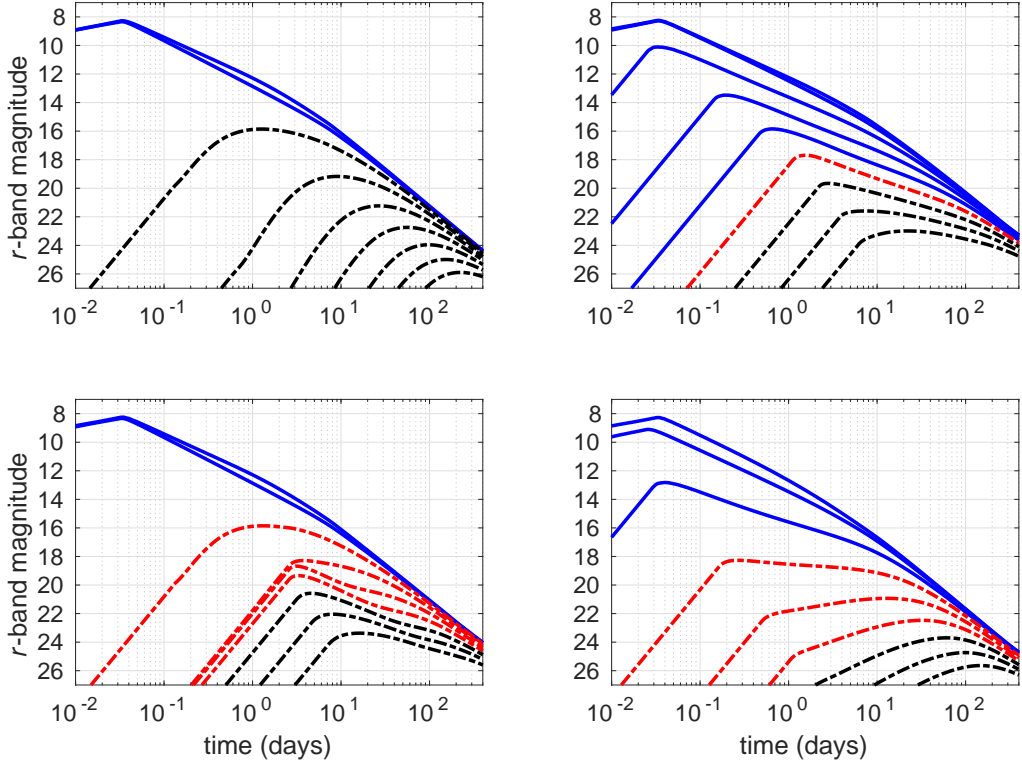}
\caption{Afterglow {\it r}-band lightcurves for jets at 200 Mpc. Lightcurves are plotted for an observer at 5$^\circ$ increments in the range $0^\circ \leq \theta_{\rm obs} \leq 40^\circ$. The model values used in each case are: (top left) $\theta_c=\theta_j=6^\circ$ for the homogeneous jet; (bottom left) $\theta_c=6^\circ$ for the two-component jet where the second component extends to $\theta_j=25^\circ$ with 5$\%$ of the core energy and Lorentz factor; (top right) $\theta_c=6^\circ$ for the power-law jet with an index $k=2$ for $\theta_c<\theta\leq25^\circ$; and (bottom right) $\theta_c=6^\circ$ for the Gaussian jet with a maximum $\theta_j=25^\circ$. Jets have an isotropic equivalent blast energy of $2\times 10^{52}$ erg, a bulk Lorentz factor $\Gamma=100$ for the core region, and an ambient medium density of $n=0.1$. Blue lines indicate the afterglow of a GRB; red dashed lines indicate an on-axis orphan afterglow i.e. within the wider jet opening angle but with suppressed prompt emission; black dashed-dotted lines indicate an off-axis orphan afterglow. Figure adapted from Lamb \& Kobayashi 2017c}
\label{figLC}
\end{figure}

An orphan afterglow population can reveal jet structure.
The characteristic lightcurves for the optical (observed $r$-band) emission from four jet structures are shown in Figure \ref{figLC}.
The lightcurves show afterglow at various inclinations for jets with a given structure:
homogeneous jets, where the jet has a uniform energy and Lorentz-factor with angle from the central axis;
two-component jets, where an energetic and fast core is surrounded by a wider sheath component with a fraction of the core energy;
power-law jets, where the energy and Lorentz-factor reduce with angle from the core edge following a negative index power-law;
and Gaussian jets, where the jet parameters follow a Gaussian distribution with angle from the central axis.
A detailed description of these models is given in \cite{2017MNRAS.472.4953L}.

Where wider jet components have a low-Lorentz factor, the prompt GRB emission is suppressed similar to a low-$\Gamma$ jet case.
In Figure \ref{figLC} the lightcurves associated with an inclination that also produces a detectable GRB are shown as blue solid lines, failed-GRB afterglows as red dashed-dotted lines, and off-axis orphan afterglows as black dashed-dotted lines.
The presence of jet structure is revealed at inclinations greater than the jet core angle.
Thus orphan afterglows from NS/BH-NS mergers can be used to indicate the presence of extended jet structure beyond the homogeneous model.

For a GW detected population of NS/BH-NS mergers, the fraction of jet EM counterparts depends on the jet model.
Where the jet population is made up of homogeneous jets with a half-opening angle $\theta_j=6^\circ$, then the fraction of the population with a jet afterglow peak flux brighter than $r$-band magnitude 21 is $\sim 13.6\%$, of this fraction $\sim 13\%$ are GRB afterglows.
Two-component jets have $\sim 30.0\%$ of the population brighter than magnitude 21, $\sim 9\%$ of these are GRB afterglows.
Power-law jets $\sim 36.9\%$ of the population brighter than magnitude 21, where $\sim 59\%$ of these are GRB afterglows.
Gaussian jets $\sim 13.3\%$ of the population brighter than magnitude 21, where $\sim 74\%$ of these are GRB afterglows.
Here a GRB is defined as being detectable by {\it Swift}/BAT if the merger occurs within the instruments field-of-view.





\section{Summary}\label{Sum}

EM counterparts from relativistic merger-jets accompanying GW detected NS/BH-NS mergers will reveal the structure and dynamical properties of short GRB jets.
If a significant population of merger-jets result in collimated low-$\Gamma$ outflows, then a hidden population of afterglow-like transients will be revealed.
Such a population can be used to constrain the Lorentz-factor distribution of a population of merger-jets.
Alternatively, the X-ray, optical or radio afterglow from a jet will reveal the presence of structure where the system is favourably inclined, i.e. $i\sim 20-40^\circ$.
Sharp lightcurve peaks, re-brighening of the afterglow during the decline after peak, or a shallow rise index pre-peak are all signatures of a structured jet viewed at an inclination greater than the core angle.

For jets inclined at angles much greater than the $\gamma$-ray bright core region an associated GRB detection is not expected.
However, the scattering of the prompt emission via a cocoon of a mergers dynamical ejecta could result in a faint GRB seen at such wide inclinations \citep{2017arXiv171100243K}.
Such a low-luminosity GRB will have an afterglow seen off-axis and peaking at $\sim100$ days.
The shape of the lightcurve for this afterglow will reveal any intrinsic jet structure.

A significant population of failed GRBs from merger-jets, or the presence of extended structure beyond a $\gamma$-ray bright jet-core, increases the rate of optical transients in an untriggered deep, $m\lesssim26$, optical survey (e.g. LSST).
These jet structures and dynamical qualities can equally be applied to a long GRB population.
For a discussion of the transient rates from such jets for optical surveys see \cite{2017arXiv171200418L}.

\section*{Acknowledgements}
GPL was able to attend the symposium due to International Astronomical Union (IAU) and Royal Astronomical Society (RAS) travel grants

\end{document}